\documentclass[10pt]{article}

\usepackage{graphicx}
\graphicspath{%
    {converted_graphics/}
    {F:/}
}
\begin{document}

\title{Schwarzschild-like exteriors for  stars in Kaluza-Klein gravity} 
\author{J. Ponce de Leon\thanks{E-mail: jpdel@ltp.upr.clu.edu; jpdel1@hotmail.com}\\ Laboratory of Theoretical Physics, Department of Physics\\ 
University of Puerto Rico, P.O. Box 23343, San Juan, \\ PR 00931, USA} 
\date{March 2010}

\maketitle

\begin{abstract}

In this work we examine the effective  four-dimensional world  that emanates from  a general class of static spherical Ricci-flat solutions in Kaluza-Klein gravity in $D$-dimensions. By means of dimensional reduction we obtain a family of asymptotically flat Schwarzschild-like metrics for which all the components of the Ricci tensor, except for $R_{11}$,  are zero. Although the reduced spacetime is not empty, it is similar to vacuum in the sense that the effective matter  satisfies an equation of state which is the generalization of $(\rho + 3p) = 0$ for ``nongravitating matter" in $4D$.
In Kaluza-Klein gravity these Schwarzschild-like metrics describe the exterior of  a spherical star without rotation.  
In this framework, we generalize the well-known Buchdahl's theorem for  
perfect fluid spheres whose mass density  does not increase outward. Without any additional assumptions, we develop the most general expression for the compactness  limit of a star. We provide some numerical values for it, which in principle are observationally testable and allow us to compare and contrast different theories and exteriors. We find that in Kaluza-Klein gravity the compactness limit of a star can be larger than $1/2$, without being a black hole: the general-relativistic upper  limit $M/R < 4/9$ is increased as we go away from the Schwarzschild vacuum exterior. We show how this limit depends on the number of dimensions of spacetime, and demonstrate  that the effects of gravity are stronger in $4D$ than in any other number of dimensions.

\end{abstract}

\medskip

PACS: 04.50.+h; 04.20.Cv

{\em Keywords:} Kaluza-Klein Theory; General Relativity; Stellar Models.

\newpage

\section{Introduction}
The concept of black holes in the context of general relativity
theory arises with the well-known Schwarzschild solution, which
according to Birkhoff's theorem is the most general spherically symmetric, asymptotically flat,  vacuum solution of the Einstein field equations. It describes the gravitational field outside an isolated spherical, non-rotating mass or black hole.  For a perfect fluid sphere with a mass density which does not increase outwards, Buchdahl \cite{Buchdahl} showed that the ratio of its gravitational mass $M$ to the coordinate radius $R$ satisfies the inequality $GM/R \leq 4 / 9$, i.e. under these conditions  no static spherically
symmetric star is possible in general relativity with a radius
less than 9/8 of the Schwarzschild radius $R_{Schw} = 2 G M$. By definition, any non-rotating and non-charged mass that is smaller than the Schwarzschild radius (horizon) constitutes  a black hole.

Nowadays
there are a number of non-equivalent theories that suggest the
existence of extra dimensions. Kaluza-Klein gravity \cite{Segre}-\cite{Wesson book} and braneworld theory \cite{Randall2}-\cite{Arkani-Hamed3} are well known examples. The study of stellar structure and stellar evolution might constitute an important approach to predict observable effects from extra dimensions. However, there is a fundamental limitation. Namely, that  Birkhoff's 
theorem is no longer valid in more than four dimensions, i.e. there is no an unique asymptotically flat  vacuum solution with spatial spherical symmetry. As a consequence,  the  effective
picture in four dimensions allows the existence of different possible non-Schwarzschild scenarios for the description of the spacetime outside of a spherical star \cite{Antoniadis}-\cite{Casadio}.

Therefore, the study of stars in these theories leads to two main questions \cite{JPdeLgr-qc/0701129}:
\begin{enumerate}
  \item 
How does the existence of stars restrict the class of admissible non-Schwarzschild exteriors?

\item How does a possible deviation from the Schwarzschild vacuum exterior can affect the star parameters?
\end{enumerate}

In a recent series of papers we have investigated different aspects of these questions. In the framework of five-dimensional Kaluza-Klein gravity, without making any assumption about the stellar structure,  we have shown that  the condition that in the weak-field limit we recover the usual Newtonian physics singles out an {\it unique} effective exterior for a spherically symmetric star \cite{JPdeLgr-qc/0701129}, \cite{JPdeLgr-qc/0703094}. On the other hand, in the framework of Randall-Sundrum II braneworld scenario, without making any assumption about the bulk, or the material medium inside the star, we have proved that for {\it any}  nonstatic spherical star, without rotation,  there are  only two possible static exteriors; these   are the Schwarzschild and the `Reissner-Nordstr{\"o}m-like' exteriors \cite{JPdeLgr-qc/0711.4415}. This is quite distinct from  the case of stars in hydrostatic equilibrium which admit a much larger family of non-Schwarzschild static exteriors. 

We have also studied   the second question mentioned above in the context of static, perfect fluid, spherical stars of {\it uniform density} \cite{JPdeLgr-qc/0701129}, \cite{JPdeL0711.0998}.
  We have shown that, in principle, the compactness limit of such a star can be larger than $1/2$, without being a black hole . 
In this work we generalize this  result to {\it any} static spherical perfect fluid star in Kaluza-Klein gravity.

Thus, here we get  rid of the unrealistic assumption of uniform density, but keep the isotropy condition\footnote{Is important to emphasize the role of isotropic pressures, because for anisotropic pressures there is no upper bound on the gravitational potential of a star. See \cite{Bowers}, \cite{Old JPdeL} and references therein.}. The generalization  is attained by using an extension, to more than four dimensions, of an inequality  originally discovered by Buchdahl \cite{Buchdahl}.  
This inequality and the matching conditions  allow us to    obtain the   compactness limit of a  star  for {\it any} given  exterior spacetime. This limit is important because it indicates how much  perfect fluid matter can be packed in a given volume, without provoking gravitational collapse. 
Here we develop the most general expression for this limit and provide some numerical values for it, which in principle are observationally testable and allow us to compare and contrast different theories and exteriors.

This paper is organized as follows. In section $2$,  we derive and analyze a general class of static, spherical vacuum solutions in Kaluza-Klein gravity in $D$-dimensions, where the metric functions are independent of the extra coordinates. They generalize some well-known solutions in the literature as the  Schwarzschild-Tangherlini spacetimes and other ones with spherical symmetry in the three usual space dimensions. In section $3$, we carry out the  dimensional reduction of the solutions. We obtain a family of asymptotically flat Schwarzschild-like metrics for which all the components of the Ricci tensor, except for $R_{11}$,  are zero. In section $4$, we use them to construct the spacetime outside of a spherical star.  
In this framework, we generalize the well-known Buchdahl's theorem for  
perfect fluid spheres whose mass density  does not increase outward. Without any additional assumptions, we show that the compactness limit of a star in Kaluza-Klein gravity can be larger than $1/2$, without being a black hole: the general-relativistic upper  limit $M/R < 4/9$ increases as we go away from the Schwarzschild vacuum exterior. Our results are consistent with our previous findings and show that, as in general relativity, all these inequalities can be saturated in the case of uniform proper density from the condition that the isotropic pressure does not become infinity at the center. In section $5$ we summarize our results. Finally, in the Appendix following our previous work \cite{JPdeL-Cruz} we 
present some general inequalities for the metric functions in the interior of a static star, which extend to $D$-dimensions those obtained by Buchdahl.

\section{Static spherical vacuum solutions in Kaluza-Klein gravity}

In five-dimensional general relativity the most general time-independent vacuum solutions, with spherical symmetry in the three usual space dimensions and a Killing vector in the fifth direction, have been found and classified by Chodos and Detweiler \cite{Chodos}.   The physical properties of these solutions with zero electric charge have been discussed by a number of authors \cite{Wesson book}, \cite{Casas}-\cite{JPdeLgr-qc/0611082}. Although they are frequently described as black holes, many researchers have noticed  that instead they  represent naked singularities because the event horizon is reduced to a singular point at the center of ordinary space \cite{Dereli}-\cite{Liu}. Truly higher dimensional extensions of the Schwarzschild black hole metric have been obtained by Tangherlini \cite{Tangherlini} and generalized by Myers and Perry \cite{Myers}, who considered spacetimes with spherical symmetry in $(D - 1)$,  rather than three spatial dimensions.  

In this work we consider static spacetimes with topology $R^1\times S^{n + 3}\times K^{m}$, where $R^1$ corresponds to the time dimension and $K^{m}$ is a $m$-dimensional manifold. The $D = (4 + n + m)$ metric is assumed to be independent of the $(n + m)$ extra coordinates and in isotropic coordinates can be written as\footnote{A more general line element is studied in \cite{EffSpacetime}.}  

\begin{equation}
\label{the metric}
dS^2 = A^2(r)dt^2 - B^2(r)\left[dr^2 + r^2 d\Omega_{(n + 2)}^2\right] - C^2(r)\delta_{a b}dy^{a}dy^{b}.
\end{equation}
Here $n$ and $m$ represent the number of ``internal" and ``external" dimensions, respectively; $d\Omega_{(2 + n)}$ is the metric on a  unit $(n + 2)$-sphere  $ (n = 0, 1, 2..)$; the coordinates along the external dimensions are denoted as $y^{a, b}$ with  $a, b = 1, 2... m$.  These spacetimes generalize the ones  discussed by  Kramer \cite{Kramer}, Davidson and Owen \cite{Davidson Owen}, Chatterjee \cite{Chatterjee}, Millward \cite{Millward}, Dereli \cite{Dereli}, Tangherlini \cite{Tangherlini}, as well as the non-rotating black holes of  and Myers and Perry \cite{Myers}.

\medskip 

{\it Notation:} Before going on, it is useful to establish our notations and conventions: Indices labeled by Latin capital letters $A$, $B$ run over the full $D = (4 + n + m)$ space; indices labeled by ${{\tilde{\mu}},  \tilde{\nu}}$ run over the $(4 + n)$ subspace and the metric is called $\gamma_{{\tilde{\mu}} \tilde{\nu}}$; the indices in the conventional spacetime, with signature is $(+, -, - ,-)$,  are labeled  $\mu, \nu$ = $(0, 1, 2, 3)$; primes denote differentiation with respect to $r$; we follow the definitions of Landau and Lifshitz \cite{Landau}; and the speed of light, as well as the gravitational constant are taken to be unity.

\medskip

The metric functions in (\ref{the metric}) are solutions of the vacuum Einstein field equations $R_{AB} = 0$. The non-vanishing equations are 

\begin{equation}
\label{R00}
\frac{A''}{A'} + (n + 1)\frac{B'}{B} + \frac{n + 2}{r} + m \frac{C'}{C} = 0,
\end{equation}

\medskip
\begin{equation}
\label{R11}
\frac{A''}{A} + (n + 2)\frac{B''}{B} + m \frac{C''}{C} + \frac{B'}{B}\left[\frac{n + 2}{r} - \frac{A'}{A} - (n + 2) \frac{B'}{B}- m \frac{C'}{C}\right] = 0,
\end{equation}

\medskip

\begin{equation} 
\label{angular part}
\frac{B''}{B} + \frac{B'}{B}\left[\frac{2n + 3}{r} + \frac{A'}{A} + n \frac{B'}{B} + m \frac{C'}{C}\right] + \frac{1}{r}\left[\frac{A'}{A} + m \frac{C'}{C}\right] = 0,
\end{equation}

\medskip

\begin{equation}
\label{y-sector}
\frac{A'}{A} + (n + 1)\frac{B'}{B} + \frac{n + 2}{r} + (m - 1)\frac{C'}{C} + \frac{C''}{C'} = 0, 
\end{equation}
which correspond to $R_{0}^{0} = 0$, $R_{1}^{1} = 0$, $R_{2}^{2} = ...  = R_{n + 2}^{n + 2} = 0$ and $R_{a}^{a} = 0$, respectively.
Now combining (\ref{R00}) and (\ref{y-sector}) we get  an equation that can be easily integrated, viz.,  $A'/A = - k\;C'/C$, where $k$ is a constant of integration. Thus, $A \propto C^{- k}$ and $B^{n + 1} \propto C^{k + 1 - m}/ r^{n + 2} C'$. Substituting into (\ref{angular part}) we obtain an equation for $C(r)$, whose solution is
\begin{equation}
\label{solution for C(r)}
C(r) = C_{0}\left(\frac{a r^{n + 1} + 1}{a r^{n + 1} - 1}\right)^{\sigma},
\end{equation}
where $a$, $C_{0}$ are constants of integration and $\sigma$ is subjected to the condition
\begin{equation}
\label{condition on sigma}
\sigma^2\left[(n + 2)k^2 - 2 k m + m(m + n + 1)\right] = n + 2,
\end{equation}
which is imposed by (\ref{R11}). The final form of the remaining metric functions is given by   
\begin{equation}
\label{A, B}
A(r) = \left(\frac{a r^{n + 1} - 1}{a r^{n + 1} + 1}\right)^{\sigma k}, \;\;\;B^{n + 1}(r) = \frac{1}{a^2 r^{2(n + 1)}}\frac{(a r^{n + 1} + 1)^{\sigma(k - m) + 1}}{(a r^{n + 1} - 1)^{\sigma(k - m) - 1}}.
\end{equation}

We will see bellow that the constant $a$ is related to the total gravitational mass. Therefore, in what follows we take $a > 0$.  Also,  without loss of generality we set $C_{0} = 1$. In the case where $D = 5$ with  $n = 0$ and $m = 1$, (\ref{condition on sigma})  reduces to $\sigma^2 (k^2 - k + 1) = 1$ which is the consistency condition in Davidson-Owen solution \cite{Davidson Owen}. 

In the above equations $k$ is an arbitrary real number, i.e.
\begin{equation}
\label{range of k}
- \infty < k < \infty,
\end{equation}
and, as a consequence of (\ref{condition on sigma}), $\sigma$ is bounded from bellow and above. Namely, 
\begin{equation}
\label{limits on sigma}
|\sigma| \leq  \sigma_{max} = \frac{n + 2}{\sqrt{m(n + 1)(n + m + 2)}}, 
\end{equation}
where the maximum value is attained at  $k = m/(n + 2)$. Thus, the range of $\sigma$ decreases, moving closer to zero,  with the increase of $m$. Therefore, for any fixed $n$ the effects  of the external extra dimensions (measured by the deviation of $\sigma$ from zero) become weaker with the increase of $m$. Also, for $m = 1$, $\sigma_{max}$ steadily decreases with the increase of $n$. However, for $m > 1$ this is not so. Indeed, in this case  $\sigma_{max}$ first decreases and  subsequently  increases with the increase of $n$.  As an illustration, let us  take 

\[ 
m = (1, 2, 5, 7, 11),
\]
then 
\[
n = 0, \;\;\;|\sigma| \leq (1.15, 0.71, 0.34, 0.25, 0.17),
\]
\[
n = 1, \;\;\;|\sigma| \leq (1.06, 0.67, 0.34, 0.25, 0.17),
\]
\[
n = 5, \;\;\;|\sigma| \leq (1.01, 0.67, 0.37, 0.29, 0.20).
\]

The behavior of the metric functions near $a r^{n + 1} \sim 1$ depends on the choice of $\sigma$ and $k$. For example, when $\sigma > 0$ we find that $B \rightarrow 0$ as $a r^{n + 1} \rightarrow 1^{+}$,  in the whole range of $k$. However, when $\sigma < 0$ the same limit gives: (i) $B \rightarrow \infty$ for $k < (m - 1)/2$, (ii) $B \rightarrow 0$ for $k > (m - 1)/2$, and (iii) $B \rightarrow 2^{2/(n + 1)}$ for $k = (m - 1)/2$.
In addition, for $\sigma > 0$, $k \geq 0$ we find that $A \rightarrow 0$, $B \rightarrow 0$, $C \rightarrow \infty$ as $a r^{n + 1} \rightarrow 1^{+}$ for any number of dimensions $n$ and $m$. Consequently,  in the allowed range of $\sigma$  and $k$, there are several  families of solutions with different geometrical and physical properties. For $n = 0$ and $m = 1$, these  have been thoroughly discussed in the literature \cite{Lake}, \cite{JPdeLgr-qc/0611082}.

A simple classification of the solutions can be obtained from the analysis of the physical radius $R(r)$ of a $(n + 2)$-sphere. In the present case it  is given by 

\begin{equation}
\label{physical radius}
R(r) = r B(r) = \frac{1}{a^{2/(n + 1)}r}\left[\frac{(a r^{n + 1} + 1)^{\sigma(k - m) + 1}}{(a r^{n + 1} - 1)^{\sigma(k - m) - 1}}\right]^{1/(n + 1)}.
\end{equation}
For $ar^{n + 1} \gg 1$, $R \approx r$, regardless of the number of dimensions and the choice of $\sigma$ and $k$. However, near $a r^{n + 1} \sim 1$
we find

\begin{eqnarray}
\label{The origin}
\lim_{a r^{n + 1} \rightarrow 1^{+}}R(r) = \left\{\begin{array}{cc}
             \;\;\; 0, \;\;\;\;\;\mbox{for} \;\;\; \sigma > 0, - \infty < k < \infty, \;\;\;\mbox{and}\;\;\; \sigma < 0, \; k > (m - 1)/2,\\
\\
\infty,\;\;\;\;\;\mbox{for} \;\;\; \sigma < 0, - \infty < k < (m - 1)/2,\;\;\;\;\;\;\;\;\;\;\;\;\;\;\;\;\;\;\;\;\;\;\;\;\;\;\;\;\;\;\;\;\;\;  \\
\\
               \left(\frac{4}{a}\right)^{1/(n + 1)}, \;\;\;\;    \mbox{for} \;\;\; \sigma = 0, \;\;\;\mbox{and} \;\;\;\sigma < 0,\; k = (m - 1)/2.\;\;\;\;\;\;\;\;\;\;\;\;
               \end{array}
              \right.
\end{eqnarray}
 For the parameters $\sigma$, $k$ in the  first expression of (\ref{The origin}) we cannot interpret (\ref{solution for C(r)})-(\ref{A, B}) as a black hole solution because the ``event horizon"\footnote{For black holes in general relativity, the event horizon is defined as the surface where the norm of the timelike Killing vector vanishes. In our case the Killing vector is just $(1, 0, 0,....0)$ so its norm vanishes where $g_{00}$ does.} $ a r^{n + 1} = 1$ occurs at $R = 0$, which by itself is a  singular point, line, plane...for $m = 0, 1, 2..$ etc.  The second expression in (\ref{The origin}) is a consequence of  the fact that $dR/dr$ vanishes  at some finite value of $r$, say $\bar{r}$. Since $R_{min} = R(\bar{r}) > 0$, the solutions with $\sigma < 0$, $k < (m - 1)/2$ can be used to generate the higher dimensional counterpart  of the wormholes  solutions discussed by Agnese {\it et al} \cite{Agnese}. 
The solutions with $\sigma = 0$ and $\sigma < 0$, $k = (m - 1)/2$ are specially simple because now $k $ and $\sigma$ are fixed. Bellow we  discuss them separately.

\paragraph{Solutions with $\sigma = 0$:} In the limit $\sigma \rightarrow 0$, from (\ref{condition on sigma}) it follows  that $\sigma^2 k^2 \rightarrow 1$. If we take  $\sigma k = 1$, then 
 
\begin{equation}
\label{A, B for sigma = 0}
A(r) =   \frac{a r^{n + 1} - 1}{a r^{n + 1} + 1}, \;\;\;B^{n + 1}(r)  =  \frac{(a r^{n + 1} + 1)^2}{a^2 r^{2(n + 1)}}.
\end{equation}
Far away from a stationary source $g_{00} \sim (1 + 2\phi)$, where $\phi$ is the Newtonian gravitational potential which goes as $- M/r^{n + 1}$. Therefore, in the present case the total gravitational mass $M$ is given by
\begin{equation}
\label{total mass}
M = \frac{2}{a}.
\end{equation}
(If we had taken $\sigma k = - 1$, we would have obtained a negative mass, viz., $M = - 2/a$). With the transformation
\begin{equation}
\label{Schw Radius}
R = \frac{\left[a r^{n + 1} + 1\right]^{2/(n + 1)}}{a^{2/(n + 1)} r},
\end{equation}
the solution becomes 
\begin{equation}
\label{Schw solution}
dS^2 = \left(1 - \frac{2 M}{R^{n + 1}}\right)dt^2 - \left(1 - \frac{2 M}{R^{n + 1}}\right)^{- 1}dR^2 - R^2 d\Omega_{(n + 2)}^2 - \delta_{a b}dy^{a}dy^{b},
\end{equation}
which, up to the innocuous $m$ flat extra dimensions,  describes  the so-called Schwarzschild-Tangherlini black holes  with spherical symmetry in $(n + 3)$ rather than three spatial dimensions. 
The radius $R_{h}$ of the horizon of the black hole is given by $R_{h}= (4/a)^{1/(n + 1)}$, which in isotropic coordinates corresponds to $a r_{h}^{n + 1} = 1$, as expected. For $n = 0$ they reduce to the conventional Schwarzschild solution of general relativity.

\paragraph{Solutions with $\sigma < 0$, $k = (m - 1)/2$:} Substituting $k = (m - 1)/2$ into (\ref{condition on sigma}) we find $\sigma = - 2/(m + 1)$. To illustrate the properties of this class of solutions let us momentarily take $n = 0$, $m = 2$. For this choice,  using (\ref{solution for C(r)})-(\ref{A, B}) we obtain 
\begin{equation} 
\label{n = 0, m = 2}
dS^2 = \left(\frac{a r + 1}{a r - 1}\right)^{2/3}dt^2 - \frac{(a r + 1)^4}{a^4 r^4}\left[dr^2 + r^2 d\Omega_{(2)}^2\right] - \left(\frac{a r - 1}{a r + 1}\right)^{4/3}\left[dy_{1}^2 + dy_{2}^2\right].
\end{equation}
The interpretation of this metric is difficult because far from the origin $g_{00} \sim [1 + (4/3 a r)]$, which implies a negative mass parameter, viz., $M = - (2/3 a)$.  However,  using a double Wick rotation $t \rightarrow i y_{1}$, $y_{1} \rightarrow i t$ we  generate a solution of the field equations with positive mass. Namely, 
\begin{equation}
\label{n = 0, m = 2 after double Wick rotation}
dS^2 = \left(\frac{a r - 1}{a r + 1}\right)^{4/3}dt^2 - \frac{(a r + 1)^4}{a^4 r^4}\left[dr^2 + r^2 d\Omega_{(2)}^2\right] - \left(\frac{a r + 1}{a r - 1}\right)^{2/3}dy_{1}^2 - \left(\frac{a r - 1}{a r + 1}\right)^{4/3}dy_{2}^2, 
\end{equation}
for which     $M = (4/3a)$. Using  the transformation of coordinates $a R = [(a r + 1)^2/a r]$ this metric becomes

\begin{equation}
\label{transformed n= 0, m= 2 solution}
dS^2 = \left(1 - \frac{3 M}{R}\right)^{2/3}dt^2 - \frac{dR^2}{(1 - 3 M/R)} - R^2 d\Omega_{(2)}^2 - \frac{dy_{1}^2}{\left(1 - 3 M/R\right)^{1/3}} - \left(1 - \frac{3 M}{R}\right)^{2/3}dy_{2}^2.
\end{equation}
We see that $g_{TT} = 0$ and $g_{RR} = - \infty$ at $R = 3 M$. However, this is {\it not} a horizon but a singularity. This may be verified by evaluating the invariant geometric scalars. For example,  the Kretschmann curvature scalar $I = R_{ABCD}R^{ABCD}$ is 
\begin{equation}
\label{I for n = 0, m = 1}
I = \frac{12 M^2 (9 R^2 - 52 R M + 76 M^2)}{(R - 3 M)^2 R^6},
\end{equation}
which is manifestly divergent at $R = 3M$. The above analysis can be extended to any $n$ and $m$. The result is that the line element  (\ref{transformed n= 0, m= 2 solution}) is a member of the family of solutions

\begin{equation}
\label{general transformed solution}
dS^2 = F^{2/(m + 1)}dt^2 - F^{- 1}dR^2 - R^2 d\Omega_{(n + 2)}^2 - {F^{(1 - m)/(1 + m)}}{dy_{1}^2} - F^{2/(m + 1)}\left[dy_{2}^2 + dy_{3}^ 2+ .....+ dy_{m}^2\right], 
\end{equation}
with\footnote{A generalization of Schwarzschild-Tangherlini's spacetimes to a chain of several Ricci-flat internal spaces has been obtained in \cite{Melnikov1}. In turn, those generalized solutions are extended  to include a massless scalar field in \cite{Melnikov2}.} 
\begin{equation}
\label{f for general transformed solution}
F = 1 - \frac{(m + 1) M}{R^{n + 1}}.
\end{equation}
These are singular at $R = \left[(m + 1)M\right]^{1/(n + 1)}$ for any $m \neq 1$. However, for $m = 1$ they yield Schwarzschild-Tangherlini spacetimes with one flat extra dimension. It is important to note that (\ref{Schw solution}) and (\ref{general transformed solution}) are the {\it only} solutions for which $g_{TT} = 0$ and $g_{RR} = - \infty$ in the same region of the spacetime.
In fact, from (\ref{A, B}) and (\ref{physical radius}) it follows  that 
\begin{equation}
\label{Bdr}
B^2(r)dr^2 = \frac{\left[a^2 r^{2(n + 1)} - 1\right]^2 \; dR^2}{\left[a^2 r^{2(n + 1)} + 2\sigma(m - k)a r^{a(n +1)} + 1\right]^2} = - g_{RR}dR^2.
\end{equation}
Thus, in general $g_{RR} \rightarrow 0$ as $a r^{(n + 1)} \rightarrow 1^{+}$, except in the case where the denominator in (\ref{Bdr}) also vanishes in this limit, i.e. for
\begin{equation}
\label{limit}
1 + \sigma(m - k) = 0.
\end{equation}
This occurs in two cases only: (i) $\sigma = 0$, $(\sigma k = 1)$ which  corresponds  to Schwarzschild-Tangherlini's spacetimes,  and (ii) $\sigma(k - m) = 1$, which by virtue of (\ref{condition on sigma}) implies $k = (m - 1)/2$ and thus generates the family (\ref{general transformed solution})-(\ref{f for general transformed solution}). 

To conclude this section, we notice that for  $k = 0$ the solution (\ref{solution for C(r)})-(\ref{A, B}) generalizes the well-known zero-dipole moment soliton of Gross and Perry \cite {Gross Perry} to any number of internal dimensions\footnote{To avoid a misunderstanding we note that the Kaluza-Klein monopole of Gross, Perry and Sorkin \cite{GPS} does not belong to the class of spacetimes investigated here.}. As an illustration, let us take 
$m = 1$, 
\begin{equation}
\label{Gross-Perry  metric for m = 1, any n}
dS^2 = dt^2 - \left(1 - \frac{1}{a r^{n + 1}}\right)^{4/(n + 1)}\left[dr^2 + r^2 d\Omega_{(n + 2)}^2\right] - \left(\frac{a r^{n + 1} + 1}{a r^{n + 1} - 1}\right)^2 dy_{1}^2.
\end{equation}
We note that here $a$ does not have to be positive. If we assume $a = - \alpha^2 < 0$ and perform  a double Wick rotation $t \rightarrow i y_{1}$, $y_{1} \rightarrow i t$, then  we obtain the metric 
\begin{equation}
\label{Shw. metric for m = 1}
dS^2 = \left(\frac{1 - \alpha^2 r^{n + 1}}{1 + \alpha^2 r^{n + 1}}\right)^2 dt^2 - \left(1 + \frac{1}{\alpha^2 r^{n + 1}}\right)^{4/(n + 1)}\left[dr^2 + r^2 d\Omega_{(n + 2)}^2\right] -  dy_{1}^2,
\end{equation}
which after a simple coordinate transformation reduces to (\ref{Schw solution}) with mass parameter $M  = 2/\alpha^2$. This suggests that, as in $5D$,   the above $(4 + n)$ metrics represent  limiting configurations that are unstable to metric perturbations \cite{JPdeLgr-qc/0611082}.

\section{Dimensional reduction}

The  class of Kaluza-Klein solutions discussed in the previous section clearly demonstrates the wealth of possible physical scenarios  in higher dimensions. In order to study their observational implications, and test possible deviations from general relativity, we have to examine the effective four-dimensional world that emanates from them. 
Thus, we consider the case where $n = 0$, which physically corresponds to  spherical symmetry in the three usual spatial dimensions. For the sake of generality,  we also discuss the dimensional reduction for $n \neq 0$. Also, at the end of this section, we discuss some similarities and differences between spacetimes with  different $n$. 

To this end, we recall that the dimensional reduction of $R_{(D)}$,  the curvature scalar associated with the metric 
\begin{equation}
\label{general metric}
dS^2 = \gamma_{\mu\nu}(x)dx^{\mu}dx^{\nu} - \sum_{i = 1}^{m}N_{i}^2(x)dy_{i}^2,
\end{equation}
can be expressed as
\begin{equation}
\label{relation between the curvature invariants}
\sqrt{|g_{(D)}|}\; R_{(D)} \propto \sqrt{|g_{(4)}^{\mbox{eff}}|}\; R_{(4)} + \mbox{other terms},
\end{equation}
where  $R_{(4)}$ is the four dimensional  curvature scalar calculated from the effective $4D$ metric tensor\footnote{For the effective action in $4D$ to contain the exact Einstein Lagrangian, i.e. to deal with a constant effective gravitational constant, $g_{\mu \nu}^{\mbox{eff}}$ should be identified with the physical metric in ordinary $4D$ spacetime \cite{Segre}, \cite{Davidson Owen}, \cite{Dolan-Duff}.}
\begin{equation}
\label{effective metric}
g_{\mu \nu}^{\mbox{eff}} = \gamma_{\mu\nu}\; \prod_{i = 1}^{m} N_{i}; 
\end{equation} 
$g_{(D)}$ and $g_{(4)}^{\mbox{eff}}$ denote the determinants of the $D$-dimensional metric (\ref{general metric}) and effective $4D$ metric (\ref{effective metric}), respectively, and the ``other terms" are proportional to the Lagrangian of an effective energy-momentum tensor (EMT) $T_{\mu\nu}$ in $4D$. If we use the fact that the metric functions satisfy the field equations $R_{A B} = 0$, then we obtain a nice expression for (\ref{relation between the curvature invariants}), viz.,  

\begin{equation}
\label{dimensional reduction  of the Ricci scalar}
\sqrt{|g_{D}|}R_{(D)} = \sqrt{|g_{(4)}^{\mbox{eff}}|}\; \left[R_{(4)} - \sum_{a = 1}^{m}\frac{\partial_{\mu}N_{a}\partial^{\mu} N_{a}}{N_{a}^2} - \frac{1}{2}\sum_{a = 1}^{m}\sum_{b = 1}^{m}\left(\frac{\partial_{\mu}N_{a}}{N_{a}}\right)\; \left(\frac{\partial^{\mu}N_{b}}{N_{b}}\right)\right].
\end{equation}
Here $D = 4 + m$ and  $m \geq 1$. It shows that the choice of the factor $\prod_{i = 1}^{m} N_{i}$ in (\ref{effective metric})  assures that the first term in (\ref{dimensional reduction  of the Ricci scalar}) yields the conventional general relativity action. 
In the case under consideration $N_{i} = C(x)$. Therefore, the physics in $4D$ can be extracted from  the four-dimensional effective action 
\begin{equation}
\label{four-dimensional action}
S_{(4)} = - \frac{1}{k_{4}}\int{d^4 x \sqrt{|g_{(4)}^{\mbox{eff}}|}}\left[R_{(4)} - m\left(1 + \frac{m}{2}\right)\frac{\partial_{\mu} C \; \partial^{\mu} C}{C^2}\right], 
\end{equation}  
where $k_{4}$ is a positive constant. From the variational principle, $\delta S_{(4)} = 0$, we get the effective equations in $4D$, viz.,
\begin{equation}
\label{effective equations in 4D}
R_{\mu\nu} - \frac{1}{2}g_{\mu\nu} R = \frac{m (m + 2)}{2 C^2}\left[\partial_{\mu}C \; \partial_{\nu}C -   \frac{1}{2} g_{\mu\nu}\partial_{\alpha} C  \; \partial^{\alpha}C\right] \equiv 8 \pi T_{\mu\nu}.
\end{equation}
Here $g_{\mu\nu} \equiv g_{\mu\nu}^{\mbox{eff}}$, and $R_{\mu\nu}$ as well as $R$ are calculated with $g_{\mu\nu}$. 

Thus, for the Kaluza-Klein metric (\ref{the metric}), (\ref{solution for C(r)}), (\ref{A, B}), with $n = 0$, the physics in $4D$ is concentrated in the effective line element 

\begin{equation}
\label{effective line element in 4D}
ds^2 = \left(\frac{a r - 1}{a r + 1}\right)^{2 \varepsilon} dt^2 - \frac{1}{a^4 r^4}\frac{\left(a r + 1\right)^{2(\varepsilon + 1)}}{\left(a r - 1 \right)^{2 (\varepsilon - 1)}} \; \left[dr^2 + r^2 d\Omega^2\right], 
\end{equation} 
with
\begin{equation}
\label{definition of varepsilon}
\varepsilon \equiv \frac{\sigma (2 k - m)}{2}.
\end{equation}
The effective $4D$ energy-momentum tensor $T_{\mu\nu}$ can be calculated by substituting (\ref{effective line element in 4D}) into the l.h.s of (\ref{effective equations in 4D}). However, it is easier  to use the r.h.s., viz.,
\begin{equation}
\label{EMT from the rhs of EFE}
8 \pi T_{0}^{0} =  \frac{m (m + 2)}{4 B^2 C^{m + 2}}\left(\frac{d C}{d r}\right)^2,  \;\;\;T_{1}^{1} = - T_{0}^{0}, \;\;\;\; T_{3}^{3} = T_{2}^{2} = T_{0}^{0}.
\end{equation}
Taking $B$ and $C$ from (\ref{solution for C(r)}), (\ref{A, B}) and using the condition (\ref{condition on sigma}), we find
\begin{equation}
\label{effective EMT in isotropic coordinates}
8\pi T_{0}^{0} = \frac{4 a^6 r^4 (1 - \varepsilon^2)(a r - 1)^{2(\varepsilon - 2)}}{(a r + 1)^{2(\varepsilon + 2)}}.
\end{equation}
It should be noted that for an observer in $4D$, who is not aware of the extra dimensions, $\varepsilon$ is a free parameter in the solution and  measures the deviation from Schwarzschild.

The line element (\ref{effective line element in 4D}) acquires a more familiar form in terms of the Schwarzschild-like coordinate $R$ defined by 
\begin{equation}
\label{transformation of coordinates for the isotropic 4D solution}
R = r\left(1 + \frac{1}{a r}\right)^2, 
\end{equation}
and setting $a = (2/\varepsilon M)$. Indeed, (\ref{effective line element in 4D}) becomes

\begin{equation}
\label{effective 4D solution in a new parameterization}
ds^2 = \left(1 - \frac{2 M/\varepsilon}{ R}\right)^{\varepsilon}\; dt^2 - \frac{dR^2}{\left(1 - \frac{2 M/\varepsilon}{ R}\right)^{\varepsilon}} - R^2 \left(1 - \frac{2 M/\varepsilon}{ R}\right)^{1 - \varepsilon}\left(d\theta^2 + \sin^2\theta d\phi^2\right).
\end{equation}
This $4D$ metric has been obtained previously by the present author  in the context of Davidson-Owen solutions \cite{Davidson Owen} (these correspond to $m = 1$). This means that the effective metric  in $4D$ is independent of the number of external extra coordinates. It ``generalizes" the Schwarzschild vacuum metric  $(\varepsilon = 1)$,  and  has been widely discussed as a possible non-Schwarzschild exterior for stellar models within the context of Kaluza-Klein gravity  \cite{JPdeLgr-qc/0701129}-\cite{JPdeLgr-qc/0703094}. We have shown that  (\ref{effective 4D solution in a new parameterization})  is  compatible with  (i) Newtonian physics, in the weak-field limit; (ii) the general-relativistic Schwarzschild limit for $\varepsilon = 1$; (iii) the dominant energy condition, and (iv) the (weak) equivalence principle, even in the non-Schwarzschild case where $\varepsilon \neq 1$. 

In terms of $R$ the EMT becomes

\begin{equation}
\label{ETM in varepsilon parameterization}
8 \pi T_{0}^{0} = \frac{(1 - \varepsilon^2)M^2}{\varepsilon^2 R^4}\left(1 - \frac{2 M/\varepsilon}{R}\right)^{(\varepsilon - 2)}, \;\;\;T_{1}^{1} = - T_{0}^{0}, \;\;\;T_{3}^{3} = T_{2}^{2} =  T_{0}^{0}. 
\end{equation}
Since $\left(T_{0}^{0} - T_{1}^{1} - T_{2}^{2} - T_{3}^{3}\right) = 0$, this spacetime is `similar' to Schwarzschild vacuum in the sense that it has no effect on gravitational interactions. However, it is more realistic because instead of being absolutely empty $T_{0}^{0} = T_{1}^{1} = T_{2}^{2} =  T_{3}^{3} = 0$, it is consistent with the existence of quantum zero-point fields \cite{WessonEssay}. It is not difficult to show that (\ref{effective 4D solution in a new parameterization}) and (\ref{ETM in varepsilon parameterization}) are equivalent to the static, spherically symmetric solution of the coupled Einstein-massless scalar field equations originally discovered by Fisher \cite{Fisher} and rediscovered by Janis, Newman and Winicour \cite{JNW}. 

\subsubsection{Dimensional reduction for $n > 0$}

Although from an observational point of view the models with $n > 0 $ might not be of prime concern, it is of theoretical interest to study how the physics depends on the number of internal dimensions. 

It is not difficult to verify that $g_{{\tilde{\mu}} \tilde{\nu}}^{\mbox{eff}}$, the effective metric in $D = (4 + n)$, is obtained             from $\gamma_{{\tilde{\mu}} \tilde{\nu}}$ as 

\begin{equation}
\label{factor for different n}
g_{{\tilde{\mu}} \tilde{\nu}}^{\mbox{eff}} = \gamma_{{\tilde{\mu}} \tilde{\nu}}\; \prod_{i = 1}^{m} N_{i}^{q(n)},\;\;\;q(n) = \frac{2}{n + 2}. 
\end{equation}
Similar to the discussion leading to (\ref{dimensional reduction  of the Ricci scalar}), with this choice the dimensional reduction of the $m$ external dimensions yields 

\begin{equation}
\label{dimensional reduction for n neq 0}
\sqrt{|g_{(D)}|}\; R_{(D)} \propto \sqrt{|g_{(4 + n)}^{\mbox{eff}}|}\; \left[R_{(4 + n)} + \mbox{constant}\times \frac{\partial_{\mu}C \; \partial^{\mu}C}{C^2}\right],
\end{equation} 
where the constants depend on the choice of $n$ and $m$. In this way, the gravitational action  has the standard form
\begin{equation}
\label{gravitational  action for any D}
S_{(D)} = - \frac{1}{k_{D}}\int{d^D x \sqrt{|g_{(D)}^{\mbox{eff}}|}}\; R_{(D)} , 
\end{equation}
in any number of dimensions.

In the case under consideration $g_{00}^{\mbox{eff}} = C^{2 m/(n + 2)} A^2$, $g_{11}^{\mbox{eff}} = - C^{2 m/(n + 2)} B^2$, $g_{22}^{\mbox{eff}} = - C^{2 m/(n + 2)} B^2 r^2$, etc. Thus the effective gravity in $4 + n$ is governed by the line element

\begin{equation}
\label{effective line element n }
ds^2 = \left(\frac{a r^{n + 1} - 1}{a r^{n + 1} + 1}\right)^{2 \varepsilon} dt^2 - \frac{1}{\left(a r^{n + 1}\right)^{4/(n + 1)}}\frac{\left(a r^{n + 1} + 1\right)^{2 (\varepsilon + 1)/(n + 1)}}{\left(a r^{n + 1} - 1 \right)^{2(\varepsilon - 1)/(n + 1)}} \; \left[dr^2 + r^2 d\Omega^2_{(n + 2)}\right], 
\end{equation} 
where 
\begin{equation}
\label{varepsilon  for n }
\varepsilon = \frac{\sigma \left[(n + 2) k - m\right]}{n + 2}.
\end{equation}
From the asymptotic behavior of the metric we get the total mass
\begin{equation}
\label{total mass for n = 2}
M = \frac{2 \varepsilon}{a}.
\end{equation}
Now, introducing the Schwarzschild-like coordinate
\begin{equation}
\label{Schw-like R for n }
R = r\left(1 + \frac{1}{a r^{n + 1}}\right)^{2/(n + 1)}
\end{equation}
the metric becomes

\begin{equation}
\label{effective line element for n in Schw-like coordinates} 
ds^2 = \left(1 - \frac{2 M/ \varepsilon}{R^{n + 1}}\right)^{\varepsilon} dt^2 -  \left(1 - \frac{2 M/\varepsilon}{R^{n + 1}}\right)^{- (n + \varepsilon)/(n + 1)}  dR^2 - R^2\left(1 - \frac{2 M/\varepsilon}{R^{n + 1}}\right)^{(1 - \varepsilon)/(n + 1)}d\Omega^2_{(n + 2)}. 
\end{equation} 

\medskip
For $n = 0$ we recover (\ref{effective 4D solution in a new parameterization}). Also,  for $n \neq 0$ and   $\varepsilon = 1$ it reduces to the Schwarzschild-Tangherlini black hole solutions (\ref{Schw solution}). For any other $\epsilon$ the effective $(4 + n)$ spacetime is not empty. In fact,  the  EMT is given by 

\medskip

\begin{equation}
\label{ETM, for any n}
8 \pi T_{0}^{0} = \frac{( n + 1)(n + 2) (1 - \varepsilon^2) M^2}{2\;  \varepsilon^2 R^{2(n + 2)}}\left(1 - \frac{2 M/\varepsilon}{R^{n + 1}}\right)^{(\varepsilon - n - 2)/(n + 1)}, \;\;\;T_{1}^{1} = - T_{0}^{0}, \;\;\;\;\;\;T_{2}^{2} = T_{3}^{3} = \cdots = T_{n + 3}^{n + 3} =  T_{0}^{0}, 
\end{equation}
which  can be interpreted as  a massless scalar field in $(4 + n)$ dimensions. For $n = 0$ it reduces to  (\ref{ETM in varepsilon parameterization}), as expected.

\subsection{Properties of the effective spacetimes}

It should be noted that the dimensional reduction eradicates the geometrical and physical differences between the three   families of higher-dimensional  solutions discussed in (\ref{The origin}). In particular, the effective $(4 + n)$ spacetime shows no evidence of the different  nature of the singularity of $g_{RR}$ near $a r^{n + 1} = 1$, which is  revealed  by (\ref{Bdr}). Thus, regardless of their specific properties, all the solutions discussed in section $2$ yield the same effective spacetime in $(4 + n)$.   

 From (\ref{ETM, for any n}) we find that  the components of the EMT satisfy  the equation 
\begin{equation}
\label{trace of EMT}
T = ( n+ 2) T_{0}^0.
\end{equation}
Substituting this into (\ref{einstein}) we find that the effective spacetimes satisfy the equations
\begin{equation}
\label{Ricci tensor for the effective metrics}
R_{0}^{0} =  R_{2}^{2} = R_{3}^{3} = \cdots = R_{n + 3}^{n + 3} =  0, \;\;\;\; R_{1}^{1} = 16 \pi T_{1}^{1}.
\end{equation}
We now recall that in the case of a constant, asymptotically flat,  gravitational field there is an expression for the total energy of matter plus field, which is an integral of $R_{0}^{0}$ over the volume $V$ occupied by the matter\footnote{In Landau and Lifshitz \cite{Landau} the discussion is in $4D$, however it can be extended to any number of dimensions} \cite{Landau}, viz.,  $M = \alpha \int{\sqrt{|g|}R_{0}^{0}dV}$, where the constant of proportionality $\alpha$ depends on the number of dimensions, e.g., $\alpha = 1/4\pi$ in $4D$. In conventional general relativity this expression  is known as the Tolman-Wittaker formula.

Since $R_{0}^{0} = 0$, it follows that the gravitational mass of any spherical shell is just zero.  This conclusion holds for any $n$ and $\varepsilon$, which include the Schwarzschild-Tangherlini black holes, as well as the familiar Schwarzschild solution of general relativity. Clearly, this is a consequence of the fact that the scalar field is {\it massless}. 

We note that 
\begin{equation}
\label{nongravitating matter}
R_{0}^{0} = \rho + p_{r} + (n + 2)p_{\perp} = 0, \;\;\; p_{r} = - T_{1}^{1}, \;\;\;p_{\perp} = - T_{2}^{2},
\end{equation}
generalizes to $n$ dimensions the well-known equation of state $\left(\rho + p_{r} + 2 p_{\perp}\right) = 0$  for nongravitating matter in $4D$, which in turn generalizes to anisotropic matter the  equation of state $(\rho + 3 p) = 0$ for a perfect fluid that has no effect on gravitational interactions \cite{Gott}-\cite{Presentauthor}. 

Finally, we note that at large distances from the origin, i.e. for   $R \gg \left(2 M\right)^{1/(n + 1)}$,  the line element (\ref{effective line element for n in Schw-like coordinates}) becomes

\begin{equation}
\label{asymptotic form}
ds^2 = ds^2_{0} - \frac{2 M}{R^{n + 1}}\left[dt^2 + \frac{(\varepsilon + n)}{\varepsilon (n + 1)}\; dR^2 + \frac{(1 - \varepsilon)}{\varepsilon (n + 1)}\; R^2\;  d \Omega_{(n + 2)}\right] + O\left(\left(\frac{M}{R^{n + 1}}\right)^2\right). 
\end{equation}
The second term represents a small correction to the Minkowski metric $ds_{0}^2$ in $(4 + n)$.  Since at large distances from  the sources every field appears centrally symmetric, it follows  that (\ref{asymptotic form}) determines the metric at large distances from any system of bodies \cite{Landau}. For $\varepsilon = 1$ the correction is identical to the one in general relativity, for any $n$.  However, this is not so for $\varepsilon \neq 1$. This could serve in astrophysical observations to detect possible deviations from general relativity.

\section{Physical interpretation}

The metric of the effective spacetime (\ref{effective line element n }) contains a singularity at $a r^{n + 1} = 1$, which, in principle, can be visible to an external observer. Only in the Schwarzschild-Tangherlini  limit  $(\varepsilon \rightarrow 1)$ it is covered by an  event horizon; for any other value of $\varepsilon$ the horizon is reduced to a singular point.  

However, the presence of naked singularities makes everybody uncomfortable and according to the cosmic censorship hypothesis they should not be realized in nature.
In order to avoid them, we have to exclude the central region and require $a r^{n + 1} > 1$. What this suggests is that the asymptotically flat metric (\ref{effective line element n }) should be used to describe  the gravitational field outside of the core of a spherical matter distribution. The ``interior" region  has to  be described by some solution of the field equations, which must be regular at the origin and  not necessarily asymptotically flat.

In this interpretation,  the effective exterior is not Ricci-flat (\ref{Ricci tensor for the effective metrics}), except for $\varepsilon = 1$. However, we have just seen that the exterior scalar field is gravitationally innocuous, in the sense that,  as in conventional general relativity, any shell outside of the source carries zero gravitational mass.

We now proceed to investigate this interpretation. We use the standard boundary conditions to study the question of how a possible deviation from the Schwarzschild vacuum exterior can affect the star parameters. Instead of restricting our discussion to  a particular equation of state  for the stellar interior, here we only assume that (1) the matter inside the star is a perfect fluid, and that (2) the energy
density is positive and does not increase outward. Under these conditions we will be able to extend the well-known Buchdahl's theorem of general relativity to stellar models in Kaluza-Klein gravity.

Thus, we assume that the spacetime outside of a spherical body is described  by the line element (\ref{effective line element for n in Schw-like coordinates}), which we now denote us\footnote{Note that  the $r$ in (\ref{exterior metric in KK gravity with n dimensions}), as well as in the rest of this section, represents what we have previously called `Schwarzschild-like coordinate', e.g.,  in (\ref{effective line element for n in Schw-like coordinates}) and (\ref{ETM, for any n}). Please, do not confuse it with the radial coordinate in isotropic coordinates (\ref{the metric}).} 

\begin{equation}
\label{exterior metric in KK gravity with n dimensions}
ds^2 = e^{\alpha(r)}dt^2 - e^{\beta(r)}dr^2 - r^2 e^{\mu(r)}\; d \Omega_{(n + 2)},  
\end{equation}
with
\begin{equation}
\label{metric terms for the exterior metric in KK gravity with n dimensions}
e^{\alpha(r)} = \left(1 - \frac{2 M/\varepsilon}{ r^{n + 1}}\right)^{\varepsilon},\;\;\;\;e^{\beta(r)} = \left(1 - \frac{2 M/\varepsilon}{ r^{n + 1}}\right)^{- (n + \varepsilon)/(n + 1)}\;\;\;e^{\mu(r)} = \left(1 - \frac{2 M/\varepsilon}{ r^{n + 1}}\right)^{(1 - \varepsilon)/( n + 1)}.
\end{equation}
The interior of a spherical star is assumed to be described by 
\begin{equation}
ds^2 = e^{\nu (R)}dt^2 - e^{\lambda(R)}dR^2 - R^2 d \Omega_{(2 + n)},
\label{general interior metric }
\end{equation}
which is the most general line element describing the interior of a spherical non-rotating star in $(4 + n)$ dimensions.  The functions $e^{\nu(R)}$ and $e^{\lambda(R)}$ are solutions of the Einstein field equations (\ref{einstein}).

 The exterior boundary of the star is a hypersurface  $\Sigma$ defined as $R = R_{b}$, and $r = r_{b}$ from inside and outside respectively. Standard matching conditions require continuity of the first and second fundamental forms at $\Sigma$ \cite{Israel}. For the metrics under consideration they demand
\begin{equation}
\label{first fundamental form KK exterior for n neq 0}
e^{\nu(R_{b})} = e^{\alpha(r_{b})}, \;\;\;R_{b} = r_{b}e^{\mu(r_{b})/2},
\end{equation}
and
\begin{eqnarray}
\label{second fund. form for KK exterior for n neq 0}
e^{- \lambda(R_{b})/2}\left(\frac{d \nu}{dR}\right)_{|_{R = R_{b}}} &=& \; e^{- \beta(r_{b})/2}\left(\frac{d \alpha}{dr}\right)_{|_{r = r_{b}}}, \nonumber \\
e^{- \lambda(R_{b})/2} &=& e^{- \beta(r_{b})/2}\left[\frac{1}{r} + \frac{1}{2}\left(\frac{d \mu}{dr}\right)\right]_{|_{r = r_{b}}} r_{b}\;  e^{\mu(r_{b})/2}.
\end{eqnarray}
Setting 
\begin{equation}
\label{Elamb}
e^{- \lambda(R)} = 1 - \frac{2 m(R)}{R^{n + 1}},
\end{equation}
from the second equation in (\ref{second fund. form for KK exterior for n neq 0}) we get
\begin{equation}
\label{m/R at the boundary for the KK exterior for general n}
\frac{m(R_{b})}{R_{b}} = \frac{1}{2}\left\{1 - \left[1 - \frac{(\varepsilon + 1)\phi_{g}}{\varepsilon}\left(\frac{R_{b}}{r_{b}}\right)^{n + 1}\right]^2 \left[1 - \frac{2 \phi_{g}}{\varepsilon}\left(\frac{R_{b}}{r_{b}}\right)^{n + 1}\right]^{- 1}\right\},
\end{equation} 
where $\phi_{g}$ is the surface gravitational potential, viz.,
\begin{equation}
\label{surface gravitational potential}
\phi_{g} \equiv \frac{M}{R_{b}^{n + 1}}.
\end{equation}
The above conditions require continuity of $T_{1}^{1}$, i.e. the ``radial" pressure,   across $\Sigma$. We find,
\begin{equation}
\label{continuity of the pressure for KK  exterior n neq 0}
8 \pi R_{b}^2 p(R_{b}) = \frac{( n + 1)(n + 2)(1 - \varepsilon^2)\phi_{g}^2}{2 \varepsilon^2}\left(\frac{R_{b}}{r_{b}}\right)^{2(n + 2)} \left[1 - \frac{2\phi_{g}}{\varepsilon}\left(\frac{R_{b}}{r_{b}}\right)^{n + 1}\right]^{(\varepsilon - n - 2)/(n + 1)}
\end{equation} 

\paragraph{Buchdahl's limit for $\varepsilon = 1$:} We note that for $\varepsilon = 1$, we recover the Schwarzschild case, i.e.,  $R_{b} = r_{b}$  and   

\begin{equation}
\label{boundary conditions for the Tangherlini  exterior }
m(R_{b}) = M, \;\;\;p(R_{b}) = 0.
\end{equation}    
Substituting this into (\ref{desired equation}), after a simple algebra we obtain
\begin{equation}
\label{surface grav. potential for Tagherlini exterior}
\frac{M}{R_{b}^{n + 1}} \leq \frac{2 (n + 2)}{(n + 3)^2}.
\end{equation}
Thus, 
\begin{equation}
\label{g00 for Tangherlini exterior}
g_{00} \geq \left(\frac{n + 1}{n + 3}\right)^2.
\end{equation} 
For $n = (0, 1, 2, 3, 4, 5, 6, 7)$, i.e. $D =  (4, 5, 6, 7, 8, 9, 10, 11)$ we find
\begin{eqnarray}
\label{grav. pot and g00 for Tangherlini solution}
\frac{M}{R_{b}^{n + 1}} &\leq& (4/9,\; 0.375,\; 0.320,\; 0.278,\; 0.245,\; 0.219,\; 0.198,\; 0.180), \nonumber \\
g_{00} &\geq& (1/9,\; 0.250,\; 0.360,\; 0.444,\; 0.510,\; 0.563,\; 0.605,\; 0.640),
\end{eqnarray}
respectively. We note that for $n = 0$, which corresponds to a matter distribution that has spherical symmetry in three spatial dimensions (rather than in $n + 3$), we recover the usual Schwarzschild values as expected.

An important observational parameter is the redshift $Z(R) = 1/\sqrt{g_{00}(R)} - 1$ of the light emitted from a point $R$ inside the sphere to infinity. For Schwarzschild-Tangherlini's's exteriors $Z_{b}$, the redshift of the light emitted from the boundary surface,  is given by
\begin{equation} 
\label{Z for Schwarzschild-Tangherlini exteriors}
Z_{b}^{max} \leq \frac{2}{n + 1}
 \end{equation}

\paragraph{Buchdahl's limit for $\varepsilon \neq 1$:}

In this case  $m(R_{b}) \neq M$ and $p(R_{b}) \neq 0$. From the second equation in (\ref{first fundamental form KK exterior for n neq 0}) we get
\begin{equation}
\label{grav. pot. as a function of (R/r) n neq 0}
\phi_{g} = \frac{\varepsilon}{2\left(R_{b}/r_{b}\right)^{n + 1}}\left[1 - \left(\frac{R_{b}}{r_{b}}\right)^{2(n + 1)/(1 - \varepsilon)}\right]. 
\end{equation}
Consequently, at the surface of a body 
\begin{equation}
\label{Enu at the boundary}
g_{00}(R_{b}) = e^{\nu(R_{b})} = \left[1 - \frac{2 \phi_{g}}{\epsilon}\left(\frac{R_{b}}{r_{b}}\right)^{n + 1} \right]^{\varepsilon} = \left(\frac{R_{b}}{r_{b}}\right)^{2 \varepsilon (n + 1)/(1 - \varepsilon)}.
\end{equation}

Substituting (\ref{m/R at the boundary for the KK exterior for general n}), (\ref{continuity of the pressure for KK  exterior n neq 0})         and (\ref{grav. pot. as a function of (R/r) n neq 0}) into (\ref{desired equation}) we obtain an inequality for $(R_{b}/r_{b})$. Unfortunately, it is very cumbersome, so we omit it here. However, it  can be solved numerically for any given value of $n$ and $0 <\varepsilon < 1$. 
Then, using the allowed values  of $(R_{b}/r_{b})$ in (\ref{grav. pot. as a function of (R/r) n neq 0}) we obtain the range of $\phi_{g}$. Here we present the solution for $n = 0, 1, 2$ and some selected values\footnote{We emphasize that for $\varepsilon > 1$ the boundary conditions do not admit solutions in the  realm of positive real numbers. } of $\varepsilon$, viz., 

\begin{equation}
\label{some selected values of epsilon}
\varepsilon = \left(0.90,\;  0.80,\;  0.75,\;  0.60\right).
\end{equation}
We also present the  redshift of the light emitted from the boundary surface $Z_{b} = Z(R_{b}) = 1/\sqrt{g_{00}(R_{b})} - 1$, which in the present case is given by
\begin{equation}
\label{Zmax}
Z _{b}= \left(\frac{R_{b}}{r_{b}}\right)^{\varepsilon (n + 1)/(\varepsilon - 1)} - 1.
\end{equation}

$\bullet$ For  $n = 0$, which corresponds to spherical symmetry in ordinary three space, the boundary conditions are satisfied for 
\begin{equation}
\label{R/r for KK exterior n = 0}
(0.882,\; 0.746,\; 0.669,\; 0.383) \leq \frac{R_{b}}{r_{b}} < 1 \;\;\;\Longrightarrow \;\;\;0 < \phi_{g} \leq (0.469,\; 0.508,\; 0.538,\; 0.777).
\end{equation}
It is important to note that the upper values of $\phi_{g}$ give the Buchdahl's limit for a star of uniform density (See equation (83) in  \cite{JPdeLgr-qc/0701129}). Using (\ref{Enu at the boundary}) and (\ref{Zmax}) we find

\[
g_{00}(R_{b}) \geq (0.104,\; 0.096,\; 0.090,\; 0.056),\;\;\; \Longrightarrow Z_{b} \leq (2.096, 2.229, 2.340, 3.219).
\]

$\bullet$ For $n = 1$, and the same values selected in (\ref{some selected values of epsilon}),  we get

\begin{equation}
\label{R/r for KK exterior n = 1}
(0.962,\; 0.915,\; 0.887,\; 0.778) \leq \frac{R_{b}}{r_{b}} < 1 \;\;\;\Longrightarrow \;\;\;0 < \phi_{g} \leq (0.383,\; 0.397,\; 0.407,\; 0.455),
\end{equation}
and 

\[
g_{00}(R_{b}) \geq (0.248,\; 0.241,\; 0.237,\; 0.222)  \;\;\;\Longrightarrow \;\;\;Z_{b} \leq (1.008, 1.035, 1.053, 1.124).
\]

$\bullet$ For $n = 2$, under the same conditions, we find

\begin{equation}
\label{R/r for KK exterior n = 2}
(0.981,\; 0.958,\; 0.944,\; 0.888) \leq \frac{R_{b}}{r_{b}} < 1 \;\;\;\Longrightarrow \;\;\;0 < \phi_{g} \leq (0.326,\; 0.329,\; 0.334,\; 0.356),
\end{equation}
and 

\[
g_{00}(R_{b}) \geq (0.355,\; 0.357,\; 0.354,\; 0.343)  \;\;\;\Longrightarrow \;\;\;Z_{b} \leq (0.679, 0.673, 0.680, 0.707).
\]
The above calculations  show how Buchdahl's limit depends on $n$ and $\varepsilon$: (i) For a fixed $n$, the upper limit of $\phi_{g}$ increases as we go away from the Schwarzschild vacuum exterior, i.e., with the increase of  $(1 - \varepsilon)$; (ii)  For a fixed $\varepsilon$, the upper limit of $\phi_{g}$ decreases with the increase of $n$, which implies that the effects of gravity are stronger in $4D$ than in any other number of dimensions.

\section{Summary and concluding remarks}

The main question under investigation here has been how a possible deviation from the Schwarzschild vacuum exterior can affect the compactness  of spherical stars in equilibrium. We have discussed this question within the context of Kaluza-Klein gravity, by using a 
general class of Ricci-flat metrics in $D = (4 + n + m)$-dimensions, namely  (\ref{solution for C(r)})-(\ref{A, B}), which generalize a number of solutions in the literature .

Following a standard technique, based on the assumption that the gravitational action has the standard form (\ref{gravitational  action for any D}) in any number of dimensions, we have reduced the $m$ external dimensions. We have seen that the reduction procedure flattens out the rich diversity of higher-dimensional solutions. The effective  metrics in $(4 + n)$  constitute  a one-parameter  family of asymptotically-flat metrics given by  (\ref{effective line element for n in Schw-like coordinates}), which contain Schwarzschild-Tangherlini's spacetimes in $(4 + n)$ dimensions for  $\varepsilon = 1$ $(m = 0, \sigma = 0)$. For any other value of $\varepsilon$ the effective spacetime is not Ricci-flat because $R_{11} \neq 0$, while all the other components of the Ricci tensor vanish identically (\ref{Ricci tensor for the effective metrics}). The fact that $R_{0}^{0} = 0$ implies that the effective (or geometrical) matter in $(4 + n)$ satisfies the  equation of state $\rho + p_{r} + (n + 2) p_{\perp} = 0$, which generalizes to $n$ dimensions the well-known equation of state  $(\rho + 3 p) = 0$ for nongravitating matter in $4D$ (gravitational or Tolman-Wittaker mass is proportional to $R_{0}^{0}$). 
Thus, for any value of $\varepsilon$, the effective spacetime is  similar to Schwarzschild vacuum in the sense that it has no effect on gravitational interactions. Consequently, we can   interpret  it  as describing the gravitational field outside of a spherical star. 

To put the discussion in perspective, let us notice that in the Randall $\&$ Sundrum braneworld scenario the concept of empty space requires only the vanishing of the Ricci scalar,  while the components of the Ricci tensor are unknown without specifying the metric in the bulk (See, e.g., \cite{JPdeLgr-qc/0711.4415}).
Here the situation is much more restricted because only $R_{1}^{1}$ is allowed to be different from zero. 
The crucial point is that Kaluza-Klein and Braneworld theories are alike in one important aspect: the effective vacuum spacetime outside of an isolated star  does not have to be Ricci-flat, as in conventional $4D$ general relativity.  

The line element (\ref{effective line element for n in Schw-like coordinates}) provides the Kaluza-Klein corrections to the Minkowski metric at large distances from any system of bodies (\ref{asymptotic form}), which could serve in astrophysical observations to detect possible deviations from general relativity. Also, it allowed us to study the question under consideration.  Namely,  
using the standard matching conditions, i.e. the continuity of the first and second fundamental forms across the boundary, as well as the generalized Buchdahl's inequality (\ref{desired equation}),   we have obtained the compactness limit          for various values of $\varepsilon$ and $n$, for any perfect fluid star with a mass density which does not increase outward.  

Our analysis shows that in Kaluza-Klein gravity the compactness limit of a star can be larger than $1/2$, without being a black hole: the general-relativistic upper  limit $M/R < 4/9$ is increased as we go away from the Schwarzschild vacuum exterior. Our results are consistent with our previous findings in \cite{JPdeLgr-qc/0701129}, \cite{JPdeL-Cruz}. They show that, as in general relativity, the compactness limit  can be saturated in the case of stars with uniform proper density from the condition that the isotropic pressure does not become infinity at the center.

It should be noted that, for any $n$, the boundary conditions require $0 < \varepsilon \leq 1$,  otherwise they have no real solutions. From (\ref{varepsilon  for n }) it follows that $\varepsilon \rightarrow 0$ as $k \rightarrow m/(n + 2)$, which according to (\ref{limits on sigma}) corresponds to the maximum value of $\sigma$. In the other extreme, 
for $\varepsilon = 1$ we recover  the Schwarzschild-Tangherlini spacetimes.  
From a physical point of view $0 < \varepsilon \leq 1$ ensures the positivity of the effective energy density  (\ref{ETM, for any n}).

 Our approach  allowed  us to determine some similarities and differences between spacetimes with different number of internal dimensions: (i) Regardless of $\varepsilon$, they satisfy similar equations, viz., (\ref{trace of EMT}), (\ref{Ricci tensor for the effective metrics}); (ii) The corrections to Minkowski metric (\ref{asymptotic form}) manifestly depend on $n$; (iii) the effects of gravity decrease with the increase of $n$.

\renewcommand{\theequation}{A-\arabic{equation}}
  \setcounter{equation}{0}  
  \section*{Appendix A: Buchdahl's inequalities in $D$-dimensions}  

In this appendix, following our previous work \cite{JPdeL-Cruz} we show how Buchdahl's inequalities can be extended to any number of internal dimensions.   
We start with the Einstein field equations in $D$ dimensions, 
\begin{equation} 
R_{A B} = 8 \pi G \left[T_{A B} - {1 \over D-2} g_{A B}
T\right],
\label{einstein}
\end{equation}
where $G$, $T_{A B}$,  and $T$ represent:   the gravitational constant; the 
energy momentum tensor in $D$-dimensions; and its trace respectively. In what follows we set $G = 1$.

We will consider  the $D$-dimensional spherically symmetric metric, given by
\begin{equation}
ds^2 = e^{\nu (R)}dt^2 - e^{\lambda(R)}dR^2 - R^2 d \Omega_{(2 + n)},
\label{metric}
\end{equation}
where $d \Omega_{(2 + n)}$ is the line element on a unit $(n + 2)$ sphere; $n = D - 4$.

Now, let us assume that the $D$-dimensional energy-momentum tensor has
the form
\begin{equation}
T_{A}^{B} = \mbox{diag}(\rho, -p, -p, ..., -p),
\label{momentum}
\end{equation}
where $\rho$ is the energy density and $p$ is the isotropic pressure. 
With this choice the field equations (\ref{einstein}) reduce to

\begin{equation}
e^{- \lambda(R)} \left [{1\over R}\left(\frac{d \lambda}{dR}\right) - {n + 1 \over R^2} \right] + {n + 1 \over
R^2} = {16 \pi \rho \over n+2} ,
\label{e-lamda}
\end{equation}

\begin{equation}
e^{- \lambda(R)} \left [{1\over R} \left(\frac{d \nu}{dR}\right) + {n + 1 \over R^2} \right] - {n + 1 \over R^2}
= {16 \pi  p \over n+2} ,
\label{e-nu}
\end{equation}

\begin{equation}
\frac{1}{2}\frac{d^2 \nu}{dR^2} + \frac{1}{4}\left(\frac{d \nu}{dR}\right)^2 - \frac{1}{4}\left(\frac{d \lambda}{dR}\right)\left(\frac{d\nu}{dR}\right) - \frac{1}{2 R}\left[\left(n + 1\right)\left(\frac{d\lambda}{dR}\right) + \left(\frac{d\nu}{dR}\right)\right]+ \frac{n + 1}{R^2}\left[e^{\lambda(R)} - 1\right] = 0.
\label{e-lanu}
\end{equation}
We note that the density and pressure satisfy the relation
\begin{equation}
\label{relation between rho and p}
\frac{d p}{dR} = - \frac{\left(\rho + p\right)}{2}\;\frac{d\nu}{dR},
\end{equation}
which is equivalent to the conservation equation $T^{B}_{\;\;A; B} = 0$.

The isotropy condition given by (\ref{e-lanu}) takes a remarkable simple form with the introduction of the following notation \cite{notation}

\begin{equation}
 e^{- \lambda} = 1 - {2 m(R) \over R^{n + 1}} = Z, \;\; 
e^{\nu(R)} = Y^{2}, \;\;  R^2 = x.
\label{appropiate notation}
\end{equation}
and 
\begin{equation}
u =  \int^x_0 {d x'\over\sqrt{Z(x')}}\;.
\label{u variable}
\end{equation}
Indeed,  (\ref{e-lanu}) reduces to 
\begin{equation}
2 {d^{2}{Y}\over du^{2}} = (n + 1)\; Y {d \over dx} \left ({m \over R^{n+3}} \right )\;.
\label{new isotropy}
\end{equation}
The function $m(R)$ can be obtained from the integration of (\ref{e-lamda}), viz.,
\begin{equation}
\label{expression for m(R)}
m(R)  = {8 \pi \over n + 2} \int_0^R \rho(\bar{R}) \bar{R}^{n + 2} d\bar{R},
\end{equation}
where the constant of integration has been set equal to zero 
to remove singularities at the origin.  Thus, 
\begin{equation}
\label{rate of change of average density}
\frac{d}{d R}\left(\frac{m}{R^{n + 3}}\right) = \frac{8 \pi \rho(R)}{(n + 2) R} - \frac{(n + 3) m(R)}{R^{n + 4}}
\end{equation}
Now, following Buchdahl we assume that the energy density is positive and does not increase outward, i.e., 

\begin{equation}
{d \rho \over dR} \leq 0.
\label{condition 2}
\end{equation}
This implies that $\rho(\bar{R}) \geq \rho(R)$ in (\ref{expression for m(R)}). Consequently, 
\[
m(R) \geq \frac{8\pi R^{n + 3} \rho(R)}{(n + 2)(n + 3)}.
\]
Substituting into (\ref{rate of change of average density}) we find 
\begin{equation}
{d \over dR} \left ({m \over R^{n+3}} \right ) \leq 0.
\label{average density}
\end{equation} 
We note that  the quantity
$m/R^{n+3}$ can be identified with the mean density of
the fluid sphere in $D$ dimensions.

Now Eq. (\ref{new isotropy}) gives

\begin{equation}
{d^{2} Y \over du ^2} \leq 0,
\label{second derivative}
\end{equation}
which means that ${dY/du}$ decreases monotonically. This in turn implies   
\begin{equation}
{d Y \over du } \leq {{Y(u) - Y(0)} \over u}.
\label{mean value theorem}
\end{equation}
Since both $Y(0)$ and $u$ are non-negative, it follows that

\begin{equation}
Y^{-1} {dY \over d u } \leq \frac{1}{u}.
\label{first derivative of Y}
\end{equation}
In terms of the original variables this equation reads

\begin{equation}
\label{basic inequality}
\left (1 - {2  m(R) \over {R}^{n + 1}} \right)^{1/2} {d \nu \over dR}
\leq 2 R \left [ \int^R_0 
\bar{R} \left (1 - {2 m(\bar{R}) \over \bar{R}^{n + 1}} \right)^{-1/2} \; d\bar{R} \right ] ^{-1}.
\end{equation}
Using the fact that the average density  decreases outward (\ref{average density}), we can evaluate 
the integral in (\ref{basic inequality}) as follows 

\begin{eqnarray}
\label{evaluation of the integral}
\int^R_0  \bar{R} \left(1 - {2  m(\bar{R}) \over \bar{R}^{n + 1}}\right)^{-1/2}\; d \bar{R}\;  \geq \; \int^R_0
 \bar{R} \left  (1 - {2 m(R) \over R^{n+3}} \bar{R}^2 \right )^{-1/2}\; d\bar{R} \nonumber \\
= {R^{n+3} \over 2 m(R)} \left [ 1- \left(1-{2 m(R) \over R^{n + 1}}\right)^{1/2} \right ].
\end{eqnarray}
On the other hand, using (\ref{e-nu}) and (\ref{appropiate notation}), $\left(d\nu/dR\right)$ can be expressed as

\begin{equation}
\label{expression for nu'}
\frac{d \nu}{dR} = \frac{2\left[(n + 2)(n + 1)  m(R)/R^{n + 1} + 8 \pi  p R^2\right]}{(n + 2) \left(1 - 2  m(R)/R^{n + 1}\right)\; R}.
\end{equation}
Now, substituting (\ref{evaluation of the integral}) and (\ref{expression for nu'}) into (\ref{basic inequality}) 
we obtain

\begin{eqnarray}
\label{desired equation}
\frac{{(n + 2)(n+1)}  m(R)/R^{n + 1} + 8\pi  p R^2}{(n+2) \left(1-{2 m(R) \over
R^{n + 1}}\right)^{1/2}} \leq {2  m(R) \over R^{n + 1}} \left
[ 1- \left(1- {2  m(R) \over R^{n + 1}}\right)^{1/2} \right ]^{-1}.
\label{Buchdahl}
\end{eqnarray}
Evaluating this expression at the outer surface of a static spherical star, and using the standard matching conditions, i.e. the continuity of the first and second fundamental forms across the boundary,  we obtain the   compactness limit of a such star  for {\it any} given  exterior spacetime. 
In particular, for the Schwarzschild exterior,  (\ref{desired equation}) leads to the well-known upper mass limit  $ M/R_{b} \leq 4/9$.


\begin{thebibliography}{99}
\bibitem{Buchdahl}{H. A. Buchdahl, {\em Phys. Rev.} 
 {\bf116} (1959) 1027.}


\bibitem{Segre}{G.C. Segr{\'{e}}, ``Physics in more than four-dimensions, another look at the Kaluza-Klein theory", in Cosmology and Elementary particles. Proceedings of the first winter school of physics. World Scientific Publishing Co. Pte. Ltd. 1989.}
\bibitem{OverduinWesson}{J. M. Overduin and  P. S. Wesson, {\em Phys.Rept.} {\bf 283} (1997) 303, arXiv:gr-qc/9805018.} 
 \bibitem{Wesson book}{P.S. Wesson, {\em Space-Time-Matter} (World Scientific Publishing Co. Pte. Ltd. 1999).} 

\bibitem{Randall2}{L. Randall and R. Sundrum, {\em Phys. Rev. Lett. } {\bf 83} (1999) 4690; arXiv:hep-th/9906064.}
 \bibitem{Arkani-Hamed1}{ N. Arkani-Hamed, S.  Dimopoulos, G. Dvali and  N. Kaloper, {\em Phys.Rev.Lett.} {\bf 84} (2000) 586, arXiv:hep-th/9907209. }
 \bibitem{Arkani-Hamed2}{N. Arkani-Hamed, S. Dimopoulos and G. Dvali, {\em Phys.Lett.} {\bf B429} (1998) 263, arXiv:hep-ph/9803315.} 
  \bibitem{Arkani-Hamed3}{N. Arkani-Hamed, S. Dimopoulos and  G. Dvali, {\em Phys.Rev.} {\bf D59} (1999) 086004, arXiv:hep-ph/9807344.} 




\bibitem{Antoniadis}{I. Antoniadis, {\em Phys. Lett.} {\bf B246} (1990) 3171.}
\bibitem{Maartens1}{R. Maartens, {\em Phys. Rev.} {\bf D62}  (2000) 084023, arXiv:hep-th/0004166.}
\bibitem{Maartens2}{Roy Maartens, Frames and Gravitomagnetism, ed. J Pascual-Sanchez et al. (World Sci., 2001) pp 93-119, arXiv:gr-qc/0101059.}
\bibitem{Dadhich1}{N. Dadhich and S.G. Gosh, {\em Phys. Lett.} {\bf B518} (2001) 1,  arXiv:hep-th/0101019.}
\bibitem{Govender}{M. Govender and N. Dadhich, {\em Phys.Lett.} {\bf B538} (2002) 233,  arXiv:hep-th/0109086.}
\bibitem{Cristiano}{C. Germani and Roy Maartens, {\em Phys. Rev.} {\bf D64}  (2001) 124010,  
arXiv:hep-th/0107011.}
\bibitem{Bruni}{M. Bruni, C. Germani and R. Maartens, {\em Phys. Rev. Lett.}
{\bf 87} (2001) 231302,    arXiv:gr-qc/0108013.}
\bibitem{Kofinas}{G. Kofinas and E. Papantonopoulos, {\em J. Cosmol. Astropart. Phys.} {\bf 12}  (2004) 11, arXiv:gr-qc/0401047.}
\bibitem{Viser}{M. Visser and D. L. Wiltshire, {\em Phys.Rev.} {\bf D67} (2003) 104004,  arXiv:hep-th/0212333.}
\bibitem{Dadhich}{N. Dadhich, R. Maartens, P. Papadopoulos and V. Rezania, {\em Phys.Lett.} {\bf B487}  (2000) 1, 
arXiv:hep-th/0003061.}
\bibitem{Bronnikov}{K.A. Bronnikov, H. Dehnen and V.N. Melnikov, {\em Phys.Rev.} {\bf D68}   (2003) 024025,  arXiv:gr-qc/0304068.}
\bibitem{Bronnikov2}{K.A. Bronnikov and S-W Kim, {\em Phys.Rev.} {\bf D67}  (2003) 064027,  arXiv:gr-qc/0212112.}
\bibitem{Casadio}{R. Casadio, A. Fabbri and L. Mazzacurati, {\em Phys.Rev.} {\bf D65} (2002) 084040,  
arXiv:gr-qc/0111072.}


\bibitem{JPdeLgr-qc/0701129}{J. Ponce de Leon, {\em Class.Quant.Grav. } {\bf 24} (2007) 1755, arXiv:gr-qc/0701129.}
 \bibitem{JPdeLgr-qc/0703094}{J. Ponce de Leon, {\em Int. J. Mod. Phys.} {\bf D18} (2009) 251, arXiv:gr-qc/0703094. } 
 \bibitem{JPdeLgr-qc/0711.4415}{J. Ponce de Leon, {\em Class.Quant.Grav.} {\bf 25} (2008) 075012, arXiv: gr-qc/0711.4415.}                  
\bibitem{JPdeL0711.0998}{J. Ponce de Leon, {\em Grav.Cosmol.} {\bf 14} (2008) 65, arXiv:0711.0998.}
  
\bibitem{Bowers}{L. Bowers and E.P.T. Liang, {\em Astrophys. J.} {\bf 188} (1974) 657.}
\bibitem{Old JPdeL}{J. Ponce de Leon, {\em Phys. Rev.} {\bf D37} (1988) 309.}

\bibitem{JPdeL-Cruz}{J. Ponce de Leon and N. Cruz, {\em Gen.Rel.Gravit.} {\bf 32} (2000) 1207, arXiv:gr-qc/0207050. }
  

\bibitem{Chodos}{A. Chodos and S. Detweiler, {\em Gen. Rel. Gravit.} {\bf 14} (1992) 879.}

\bibitem{Casas}{J.A. Casas, C.P. Martin and A.H. Vozmediano, {\em Phys. Lett.} {\bf B 186} (1987) 29. }
\bibitem{Gurin}{V.S. Gurin and A.P. Trofimenko, {\em Phys. Lett.} {\bf B 241} (1990) 328.}
\bibitem{previous work}{P.S. Wesson and J. Ponce de Leon, {\em Class. Quantum Grav.} {\bf 11} (1994) 1341.}
\bibitem{LOW}{P. Lim, J. Overduin and P.S. Wesson, {\em J. Math. Phys.} {\bf 36} (1995) 6907.}
\bibitem{Agnese}{A.G.Agnese, A.P. Billyard, H. Liu and P.S. Wesson, {\em Gen. Rel. Gravit.} {\bf 31} (1999) 527.}
\bibitem{Sajko}{W.N. Sajko and P.S. Wesson, {\em Gen. Rel. Gravit.} {\bf 32} (2000) 1381.}
\bibitem{Liu-Overduin}{H. Liu and J. Overduin, {\em Astrophys. J.} {\bf 538} (2000) 386, arXiv:gr-qc/0003034.}
\bibitem{Overduin}{J. Overduin, {\em Phys. Rev.} {\bf D62} (2000) 102001, arXiv:gr-qc/0007047.}
\bibitem{Lake}{K. Lake, {\em Class. Quant. Grav.} {\bf 23} (2006) 5876,   arXiv:gr-qc/0606005.}
 \bibitem{JPdeLgr-qc/0611082}{J. Ponce de Leon, {\em Int.J.Mod.Phys.} {\bf D17} (2008) 237, arXiv:gr-qc/0611082. }


\bibitem{Dereli}{T. Dereli, {\em Phys. Lett.} {\bf B161} (1985) 307.}
\bibitem{Sokolowski}{L. Sokolowski and B. Carr, {\em Phys. Lett.} {\bf B 176} (1986) 334. }
\bibitem{Chatterjee}{S. Chatterjee, {\em Astron. Astrophys.} {\bf 230} (1990) 1.}
\bibitem{Liu}{H. Liu, {\em Gen. Rel. Gravit.} {\bf 23} (1991) 759.}

\bibitem{Tangherlini}{F. R. Tangherlini, {\em Nuovo Cimento} {\bf 27} (1963) 636.}
\bibitem{Myers}{R.C. Myers and M.J. Perry, {\em Annals of Physics} {\bf 172} (1986) 304.}

\bibitem{EffSpacetime}{J. Ponce de Leon, {\em Grav. Cosmol.} {\bf 15} (2009) 345, arXiv:0905.2010.}

\bibitem{Kramer}{D. Kramer, {\em Acta Phys. Polon.} {\bf B2} (1970) 807.}
\bibitem{Davidson Owen}{A. Davidson and D. Owen, {\em Phys. Lett.} {\bf B 155} (1985) 247.}
\bibitem{Millward}{R.S. Millward, ``A five-dimensional Schwarzschild-like solution", arXiv:gr-qc/0603132.}

\bibitem{Landau}{L.D. Landau and E.M. Lifshitz, {\em The Classical Theory of Fields}, Fourth Edition (Butterworth-Heinemann, 2002).}

\bibitem{Melnikov1}{S.B. Fadeev, V.D. Ivashchuk and V.N. Melnikov, {\em Phys. Lett.} {\bf A 161} (1991) 98.}
\bibitem{Melnikov2}{V.D. Ivashchuk and V.N. Melnikov, {\em Grav. Cosmol.} {\bf 1} (1995) 133, arXiv:hep-th/9503223.}


\bibitem{Gross Perry}{D.J. Gross and M.J. Perry, {\em Nucl. Phys.} {\bf B226} (1983) 29.}
\bibitem{GPS}{D.J. Gross, M.J. Perry and R.D. Sorkin, {\em Phys. Rev. Lett.} {\bf 51} (1983)) 87.} 

\bibitem{Dolan-Duff}{L. Dolan and M.J. Duff, {\em Phys. Rev. Lett.} {\bf 52} (1984) 14.}
\bibitem{WessonEssay}{P.S. Wesson, {\em Phys. Essays, Orion} {\bf 5} (1992) 591.}
\bibitem{Fisher}{Z. Fisher, {\em Zh. Eksp. Teor. Fiz.} {\bf 18} (1948) 636 (in Russian), arXiv:gr-qc/9911008.}
\bibitem{JNW}{A.I. Janis, E.T. Newman and J. Winicour, {\em Phys. Rev. Lett.} {\bf 20} (1968) 878.}


\bibitem{Gott}{J.R. Gott and M.J. Rees, {\em Mon. Not. R. Astron. Soc.} {\bf 227} (1987) 453.}
\bibitem{Kolb}{E. Kolb, {\em Astrophys J.} {\bf 344} (1989) 543.}
\bibitem{Presentauthor}{J. Ponce de Leon, {\em Gen. Rel. and Gravit.} {\bf 11} (1993) 1123.}
\bibitem{Israel}{W. Israel, {\em Nuovo Cimento} {\bf B44}, (1966) 1;[Erratum-ibid. {\bf B48} (1967) 463].}
\bibitem{notation}{J. Ponce de Leon, {\em J. Math. Phys.} {\bf 29} (1988) 197.}



\end{thebibliography}
\end{document}